\documentclass[twocolumn,nofootinbib,amsmath,amssymb,a4paper]{revtex4}

\usepackage{graphicx}% Include figure files
\usepackage{dcolumn}% Align table columns on decimal point
\usepackage{bm}

\input babarsym
\def\dstarelnubulo{\ensuremath{\Dstarm\ell^+\nu_{\ell}}\xspace}
\def\im{\ensuremath{\mathrm{Im}}\xspace}
\def\re{\ensuremath{\mathrm{Re}}\xspace}
\newcommand{\z}{\ensuremath{{\mathsf z}}\xspace}
\def\dM{\ensuremath{\Delta m}\xspace}
\def\dG{\ensuremath{\Delta\Gamma}\xspace}
\def\dt{\ensuremath{\Delta t}\xspace}
\def\AT{\ensuremath{A_{CP/T}}\xspace}
\def\ACPT{\ensuremath{A_{CPT}}\xspace}
\def\Imz{\ensuremath{\im \z}\xspace}
\def\dGRez{\ensuremath{\Delta\Gamma \cdot \re \z}\xspace}
\newcommand{\ImzZero}{\ensuremath{\rm Im\, \z_0}}
\newcommand{\ImzOne}{\ensuremath{\rm Im\, \z_1}}
\newcommand{\dGRezZero}{\ensuremath{\dG \cdot \re \z_0}}
\newcommand{\dGRezOne}{\ensuremath{\dG \cdot \re \z_1}}
\def\mnu{\ensuremath{M_\nu^2}\xspace}

\def\psoft{\ensuremath{\pi_s}\xspace}
\begin{document}

\title{Searches for \boldmath{$CP$}, \boldmath{$T$} and \boldmath{$CPT$} 
violation in \boldmath{\BzBzb} mixing at \babar}

\author{R. Covarelli}
 \email{covarell@slac.stanford.edu}
\affiliation{%
Universit\`{a} degli Studi di Perugia, I-06123 Perugia, Italy\\
}%

\begin{abstract}
We describe measurements of $CP$ and $CPT$ violation parameters in \BzBzb
oscillations, performed at \babar\ by using \BB\ events collected at 
the \FourS\ mass peak. These are obtained from two different 
$B$-reconstruction methods, namely: searching for two inclusive semileptonic 
\Bz\ decays in an event or partially reconstructing one of the \Bz\ mesons
in the \dstarelnubulo\ channel.
% or fully reconstructing it in hadronic channels. 
The results are in agreement with one another and with the
most recent theoretical calculations in the Standard Model scenario.  
\end{abstract}

\maketitle

\section{Theoretical overview}
$CP$ violation due to the neutral-$B$ meson oscillation processes can be
parameterized in two different scenarios: either in the Standard Model (SM)
field theory, with some extensions allowed within the
$CPT$-conservation hypothesis, or, more generally, also taking into account
the possibility of $CPT$ violation.

In the SM case, \BzBzb\ mixing is accounted for using a time-dependent
perturbation theory that leads to an effective hamiltonian
$\mathbf H_{eff} = \mathbf M - \frac{i}{2} \mathbf \Gamma$, that is the 
$2 \times 2$ projection of the actual interaction operator on the \Bz\ and
\Bzb\ states only.
In solving the eigenvalue problem, linear relations are found
connecting the $B$ flavor eigenstates to the physical states:
\begin{eqnarray}
|B_L\rangle&=&p|\Bz\rangle +q|\Bzb\rangle\nonumber ~ ,\\
|B_H\rangle&=&p|\Bz\rangle -q|\Bzb\rangle ~ .
\label{eq:mass_eigenstates}
\end{eqnarray}
In the Standard Model, the magnitude of the ratio $q/p$ is very nearly unity:
\begin{equation}
\left|\frac qp\right|^2\approx 1-\im \frac{\Gamma_{12}}{M_{12}},
\end{equation}
not just because $|\Gamma_{12}|$ is small, but
because the \CP-violating quantity $\im (\Gamma_{12}/M_{12})$
is suppressed by an additional factor
$(m_c^2-m_u^2)/m_b^2\approx 0.1$ relative to $|\Gamma_{12}/M_{12}|$.
When the remaining factors are included, the expectation is 
\mbox{$|\im ({\Gamma_{12}}/{M_{12}})|<10^{-3}$}~\cite{ciuch,bene}.

The parameter $|q/p|$ is sensitive to New Physics (NP) within a 
$CPT$-conserving theory~\cite{laplace}. Assuming 3-generation unitarity of 
the CKM  matrix~\cite{bib:KM} and tree-level processes dominated by the SM,
the \CP\ asymmetry can be modified as follows:
\begin{eqnarray}
&& \AT \simeq 2(1-|q/p|) =  \\ 
\nonumber
&& -\mathrm{Re}\left(\frac{\Gamma_{12}}{M_{12}}\right)^
{\mathrm{SM}} \frac{\sin{2 \phi_d}}{C_d^2} + \mathrm{Im}\left(\frac
{\Gamma_{12}}{M_{12}}\right)^{\mathrm{SM}} \frac{\cos{2 \phi_d}}{C_d^2} ~,
\label{eq:npasl}
\end{eqnarray} 
$C_d$ and $\phi_d$ being a generic NP amplitude and phase. Within
the NP parameter space, $|q/p| - 1$ could be enhanced up to an order of
magnitude.

If $CPT$ violation is also allowed, its effects are usually parameterized 
in terms of an additional complex parameter \z\ in the eigenstate definitions 
(\ref{eq:mass_eigenstates}):
\begin{eqnarray}
|B_L\rangle&=&p\sqrt{1-\z}|\Bz\rangle +q\sqrt{1+\z}|\Bzb\rangle\nonumber ~ ,\\
|B_H\rangle&=&p\sqrt{1+\z}|\Bz\rangle -q\sqrt{1-\z}|\Bzb\rangle ~ .
\label{eq:mass_eigenstates_cpt}
\end{eqnarray}
Kosteleck\`{y} \cite{koste} first pointed out that such parameterization
using a constant value of \z\ is not completely satisfactory, since $CPT$ 
violation in a field theory implies that Lorentz covariance, too, is 
violated. A more appropriate parameterization is achieved by setting 
$\z \propto \beta^{\mu}\Delta a_{\mu}$, where $ \beta^{\mu} \equiv 
\gamma_B(1,\vec{\beta}_B)$ is the decaying $B$-meson four-velocity
and $\Delta a_{\mu}$ is a four-vector containing NP $CPT$-violating 
coefficients. We approximate the 4-velocity of each $B$ meson by the 
\FourS 4-velocity so that \z is common to each $B$ in an event; in this way
the 3-velocity direction becomes a function of the Earth position with
respect to the Universe at the time of the \FourS\ decay, i.e.\ of the
event's sidereal time. 
\z\ is then expressed as:
\begin{equation}
\z \equiv \z(\hat{t}) = \z_0 + \z_1\cos{(\Omega\hat{t} + \phi)}
\end{equation}
where $\Omega = 2 \pi$ rad/sidereal day is the Earth rotation frequency
(1 sidereal day $\sim$ 0.997 solar days), $ \hat{t}$ is the sidereal
time, $\z_0$ and $\z_1$ contain NP coefficients.

All the analyses presented in the following exploit the $\Delta t$ dependence
of the $B$-pair decay rates in order to extract $CP$ and $CPT$ violation 
parameters, $\Delta t$ being the difference between
the decay times of the two mesons. The general form is:
\begin{eqnarray}
{{\rm d}N\over {\rm d} (\Delta t)} &\propto & e^{-\Gamma |\Delta t|}\bigg[ {1\over 2} 
c_+\cosh \left(\frac{\dG \Delta t}{2}\right) +{1\over 2}c_-\cos(\dM t) 
\nonumber \\
& &  -\re s\, \sinh \left(\frac{\dG \Delta t}{2}\right) 
+\im s \,\sin(\dM t) \bigg]. ~~~~~~\label{eq:pref2}
\end{eqnarray}
In this equation \dM\ (\dG) is the mass (lifetime) difference between the 
\Bz\ physical states, while $c_+$, $c_-$ and $s$ depend on the physical
parameters $|q/p|$ and \z, on the amplitudes of the particular processes under
consideration, and the flavor of the $B$ mesons at the time of decay. 
  
\section{The inclusive dilepton method}

Inclusive dilepton events, where both $B$ mesons decay
semileptonically  ($b \to X\ell\nu$, with $\ell=e$ or $\mu$),
represent 4\% of all \upsbb
decays and  provide a very large sample with which to study \T, \CP
and \CPT violation.
In the direct semileptonic neutral $B$ decay,
the flavor $\Bz\,(\Bzb)$  is tagged by the charge of the lepton 
$\ellp\,(\ellm)$.
This study~\cite{yeche} is performed using a
\babar~\cite{babarnim} data sample equivalent to an integrated luminosity of 
211 \invfb recorded at the \FourS resonance (232 million \BB\ pairs).

Electrons and muons are selected using information from the tracking systems,
the electromagnetic calorimeter and the muon chambers. The track momentum is 
required to be between
0.8 and 2.3 \gevc\ and certain dilepton mass ranges are vetoed in order to 
reject photon conversions, \jpsi\ and $\psi (2S)$ decays.
The separation between {\it direct} leptons $(b \to \ell)$ and background from
the $b\rightarrow c\rightarrow \ell$ decay chain ({\it cascade} leptons)
is achieved with a neural network that combines five discriminating variables:
the momenta and opening angle of the two lepton candidates,
and the total visible energy and missing momentum of the event.

Since $|q/p|$ is expected to be small, we have
determined the possible charge asymmetries induced by
charge-dependent differences in the
reconstruction and identification of electrons.
The charge asymmetry of track reconstruction is measured in the data
by comparing tracks reconstructed using only the silicon tracker with those 
passing the dilepton track selection,
The lepton identification efficiencies are measured
with a control sample of
radiative Bhabha events for electrons.
Based on tracking and identification, we fix
the charge asymmetry for the direct electrons to
$a_{e}^{dir}=1.2 \times 10^{-3}$.
In the likelihood fit, we float the
charge asymmetries $a_{\mu}^{dir}$ and $a_{\mu}^{casc}$ for direct 
and cascade muons.

\begin{figure}[hbtp]
\begin{center}
\includegraphics[width=0.3\textwidth]{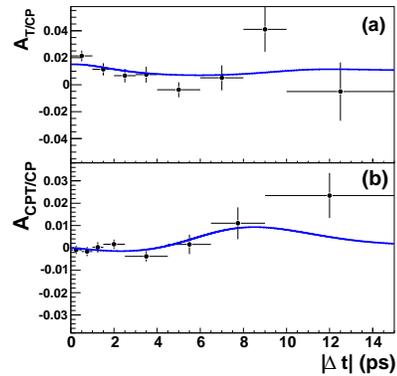}
\end{center}
\caption{ (a) $\AT$ asymmetry between $(\ell^+,\ell^+)$ and $(\ell^-,\ell^-)$.
 (b)  $\ACPT$ asymmetry between  $(\ell^+,\ell^-)$
dileptons with $\dt>0$ and $\dt<0$.}
\label{fig:AsyData}
\end{figure}

We model the contributions to our sample from \BB\ decays using
different categories of events each represented by a probability density
function (PDF) in $\dt$. 
The five categories are the following.
First, the pure signal events  with two direct leptons are
81\% of the \BB\ events.
Then, we consider two categories of cascade decays: those in which the
direct lepton and the cascade lepton come from different $B$ decays, and those 
in which the direct lepton and the cascade lepton stem
from the same $B$ decay. In addition,
a small fraction of the dilepton events originate from the decay chain 
$b \to \tau^- \to \ell^-$
which tags the $B$ flavor correctly. Finally, the remaining events consist 
mainly of leptons from the decay of a charmonium 
resonance.
The last component of the dilepton sample originates from non-\BB\ events,
and has been estimated using off-resonance data.
Figure~\ref{fig:AsyData} shows the result of the maximum likelihood fit in
terms of the projected asymmetry
 $\AT$ between $(\ell^+,\ell^+)$ and $(\ell^-,\ell^-)$
dileptons and $\ACPT$ between  $(\ell^+,\ell^-)$
dileptons with $\dt>0$ and $\dt<0$.

For $|q/p|$, the most important systematic uncertainties are due to the 
correction of electron charge asymmetries, while the dominant systematic
 uncertainties on \z\ are the imperfect knowledge of the absolute scale of the
detector and the residual uncertainties in the tracker local
alignment.

The results for \CPT and \CP violation parameters are
\begin{eqnarray*}
|q/p| -1 & = & (-0.8 \pm 2.7_{\rm (stat.)}  \pm 1.9_{\rm(syst.)})\times 10^{-3},\\
\Imz  & = & (-13.9  \pm 7.3_{\rm (stat.)} \pm 3.2_{\rm(syst.)})\times 10^{-3},\\
\dGRez & = & (-7.1 \pm 3.9_{\rm (stat.)} \pm 2.0_{\rm(syst.)})\times 10^{-3}\ps^{-1}.
\end{eqnarray*}
where \dGRez has been determined with the approximation 
$\re \z \sinh(\Delta\Gamma \dt /2) \sim \re \z \cdot \dG \cdot \dt /2$ and 
\z\ is considered as a constant parameter.

\begin{figure}[hbtp]
\begin{center}
\includegraphics[width=0.35\textwidth]{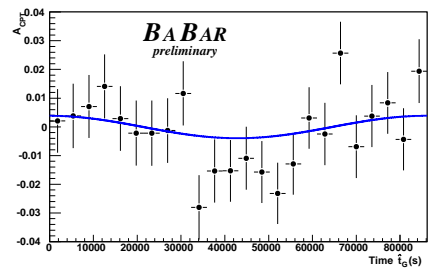}
\end{center}
\caption{The opposite-sign lepton asymmetry $A_{\CPT}$ as a 
         function of GMST (in seconds) folded over a 
         period of 24 sidereal hours. The curve is a 
         projection from the two-dimensional likelihood 
         fit onto the sidereal time axis.}
\label{fig:AsySid}
\end{figure}

A similar \dt\ fit is then performed, taking the sidereal-time dependence
into account. We use the latitude and longitude of the 
\babar\ detector, together with its orientation to determine the Lorentz 
boost direction of the \FourS in the Earth frame. 
We convert the time recording of each event occurrence to calculate GMST 
(Greenwich Mean Sidereal Time) as specified by the U.S.\ Naval 
Observatory. Figure~\ref{fig:AsySid} shows the $A_{\CPT}$ asymmetry, 
integrated over the \dt\ variable, as a function of the GMST. 

The results for sidereal-time dependent \CPT violation parameters are
\begin{eqnarray*}
\ImzZero   & = & (-14.1  \pm 7.3_{\rm (stat.)} \pm 2.4_{\rm (syst.)})\times 10^{-3},\\
\dGRezZero & = & (-7.2   \pm 4.1_{\rm (stat.)} \pm 2.1_{\rm (syst.)})\times 10^{-3}\ps^{-1},\\
\ImzOne    & = & (-24.0   \pm 10.7_{\rm (stat.)} \pm 5.9_{\rm (syst.)})\times 10^{-3},\\
\dGRezOne  & = & (-18.8   \pm 5.5_{\rm (stat.)} \pm 4.0_{\rm (syst.)})\times 10^{-3}\ps^{-1}.
\end{eqnarray*}

\section{The partial reconstruction method}

In a complementary approach \cite{nostra} the partial reconstruction of the
$\Bz \rightarrow \dstarelnubulo$ decay of one of the neutral $B$ mesons is
exploited. 
Though the available statistics is not 
as high as in the dilepton case, we can keep the charged $B$ background  
at a lower level. At the same time, since the 
reconstructed and tag side are well defined, a procedure to determine 
particle detection asymmetries from data can be carried out. In this way, no
dedicated control samples are needed to determine the asymmetry induced by the 
experimental cuts.
A \babar\ data sample equivalent to an integrated luminosity of 201 \invfb 
at the \FourS resonance (220 million \BB\ pairs) is used for this study.

Preliminary cuts for lepton selection and non-\BB\ background rejection are
similar to the inclusive dilepton analysis. Only the charged lepton from 
the \Bz\ decay and the soft pion
($\pi^-_s$) from the \Dstarm decay are reconstructed. The \Dzb\ decay is 
not reconstructed, resulting in high selection efficiency.
Due to the limited phase space available in the \Dstar decay,
we approximate the direction of the \Dstar to be that of the \psoft
and estimate the energy  $\tilde{E}_{D^{*}}$ of the \Dstar
as a linear function of the energy of the \psoft. 
We define the square of the missing neutrino mass as:
\begin{equation}
\label{eq:m2}
\mnu = \left( \sqrt{s}/2 - \tilde{E}_{\Dstar} - E_{\ell} \right)^2 -
(\tilde{\bf{p}}_{\Dstar} + {\bf{p}}_{\ell} )^2 ,
\end{equation}
where all quantities are defined in the \FourS\ frame. We neglect 
the momentum of the \Bz\ (approximately 0.34 GeV/$c$),
and identify the \Bz\ energy with half the total energy of the events 
($\sqrt{s}/2$).
$E_{\ell}$ and ${\bf{p}}_{\ell}$ are the energy and momentum vector 
of the lepton and
$\tilde{\bf{p}}_{\Dstar}$ is the estimated momentum vector of the \Dstar.
The distribution of $\mnu$ is peaked for signal events, while it is 
spread over a wide range for background events.
Backgrounds are also suppressed using cuts on the lepton momenta, 
on the soft pion momemtum, on the vertex fit probability, on
$|\Delta t|$ and $\sigma_{\Delta t}$. 

\begin{figure}[hbtp]
\begin{center}
\includegraphics[width=0.35\textwidth]{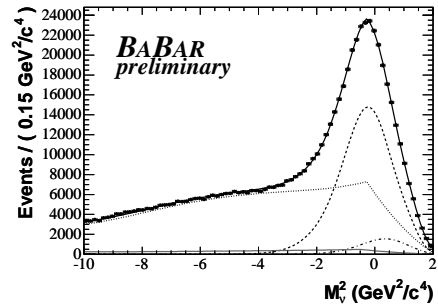}
\end{center}
\caption{Squared neutrino mass distributions for the tagged data events. 
The following 
fitted contributions are superimposed: continuum (solid grey line), 
combinatorial (dotted), $B^{\pm}$ peaking (dash-dotted), signal (dashed),
all (solid black).}
\label{fig:mnucomp}
\end{figure}

These selection criteria accept 470,877 events in the data 
sample ({\it tagged} events),
and 5,291,868 partially reconstructed events that fail only the
requirements of lepton tagging ({\it untagged} 
events). 
Fig.~\ref{fig:mnucomp} shows
the distributions of the squared neutrino mass for tagged events. 
Different components are recognized in the total sample:
{\it signal} events, due to \dstarelnubulo\ events; {\it peaking} background events, 
that correspond to decays of 
charged $B$ mesons which peak in the \mnu\ distribution, 
like $B^+ \to D^{\ast\ast 0} \ell^+ \nu_{\ell}$, $
D^{\ast\ast 0} \to D^{*-} \pi^+$;
{\it combinatorial} events, corresponding to all $\BB$ decays not included 
in the previous categories; {\it continuum} events, coming from lepton-pair
and light quark decays.
Fractions of the different components in data are determined by means of 
fits to the \mnu\ distribution: we obtain a signal fraction of
$(46.6 \pm 0.9)\%$ in the 
selected mass region  $-4.0 < \mnu < 2.0~\mathrm{GeV}^2/c^4$.

As in the dilepton case, the contributions to our sample are modeled using
probability density
functions (PDF) in $\dt$, based on the signal-background categories defined
above and on the true origin of the tagging lepton.

An original method is used to determine possible charge detection asymmetries
in the sample. We parameterize the total number 
of expected events for each tagged
and untagged category in a likelihood function, which 
contains a term coming from the shapes of the PDFs for tagged events,
an extended term for the number of tagged events and an extended term for 
the number of untagged events. 
This method 
reduces the correlation between $CP$
mixing asymmetry and particle detection asymmetry.

\begin{figure}[hbtp]
\begin{center}
\includegraphics[width=0.35\textwidth]{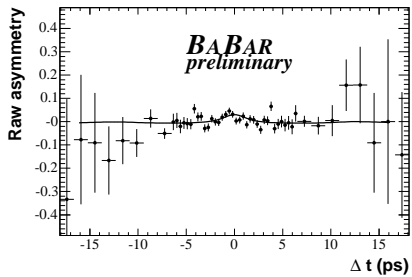}
\end{center}
\caption{Raw asymmetry between mixed positive $(s_r = 1, s_t = 1)$ and
mixed negative $(s_r = -1, s_t = -1)$ events with the PDF asymmetry
derived from the fit superimposed.
The asymmetry at low values of $|\Delta t|$ corresponds to the kaon-mistag
contribution in decay-side tagged events.}
\label{fig:asymdata}
\end{figure}

Figure~\ref{fig:asymdata} shows the result of the maximum likelihood fit in
terms of the projected asymmetries
$\AT$ between mixed positive and negative events, that we convert
in a value for $|q/p|$. The main systematic uncertainties 
comes from the \BB\ background and continuum charge asymmetries
and the signal fraction determination from the \mnu\ fit. The preliminary
result is:
\begin{equation}
|q/p| -1 = (6.5 \pm 3.4_{(\mathrm{stat.})} \pm 2.0_{(\mathrm{syst.})}) \times 
10^{-3}, 
\end{equation}
which is compatible with the absence of $CP$ violation in mixing at the
9.8\% C.L.

\section{Conclusions}
We have presented determinations of the parameters $|q/p|$, that is 
related to $CP$ violation in \BzBzb\ mixing, 
and \Imz, \dGRez, \ImzOne\ and \dGRezOne, that measure
$CPT$ violation in a theory that takes into account effects of 
the implied Lorentz violation. 

We exploit fits
to $\Delta t$, the time difference between the two $B$ decays in
two complementary approaches. In the former, two high-momentum
leptons are searched for in order to select inclusive semileptonic
\Bz\ decays.
In the latter, one of the $B$ mesons is partially reconstructed in the semileptonic 
channel \dstarelnubulo, i.e. only the lepton and the soft pion from
$\Dstarm \rightarrow \Dzb \pi^-$ decay are reconstructed, 
while the flavor of the 
other $B$ is determined by means of lepton tagging.    
We use \BzBzb\ pair events,
collected by the \babar\ detector in the period
1999-2004.

\begin{figure}[hbtp]
\begin{center}
\includegraphics[width=0.25\textwidth]{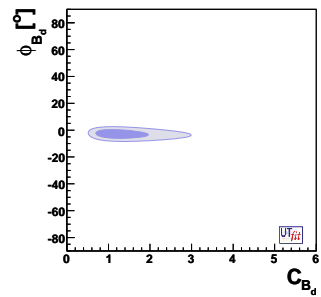}
\end{center}
\caption{Allowed regions for the NP parameters $C_d$ and $\phi_d$ at the
68\% (dark blue) and 95\% (light blue) confidence levels.}
\label{fig:utfit}
\end{figure}

No evidence is found of $CP$ violation in mixing with either of the two 
methods: both results are compatible with the SM expectations and
with previously published \babar\ results~\cite{fmv,bozzi}. The experimental
values of $\AT$ are used as constraints~\cite{utfit}, together with other measurements
related to the unitarity triangle, in a global fit to the
CKM theory parameters that also 
includes the possibility of NP. Figure~\ref{fig:utfit} shows the 
most recent results of this fit in terms of the allowed regions for the
parameters $C_d$ and $\phi_d$ defined in Eq.~(\ref{eq:npasl}),
corresponding to a 95\% confidence interval of $[0.59,2.51]$ for $C_d$
and $[-7.0,1.0]^\circ$ for $\phi_d$.
  
\begin{figure}[hbtp]
\begin{center}
\includegraphics[width=0.35\textwidth]{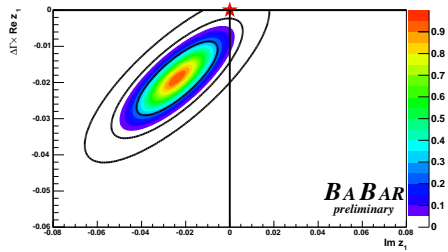}
\end{center}
\caption{Allowed regions for the $CPT$-violating  
parameters \ImzOne\ and \dGRezOne\ at
various confidence levels. The red star represents the SM
expectation and the solid black ellipses correspond to
1, 2 and 3$\sigma$ significances.}
\label{fig:CPTv}
\end{figure}

For the \Imz, \dGRez\ pair, the results are compatible with the SM
expectation (0,0) within 1.5$\sigma$. When sidereal-time
dependence is taken into account, the result in the \ImzOne-\dGRezOne
plane has a significance of 2.2$\sigma$, compatible with the
absence of $CPT$ violation in mixing (Figure~\ref{fig:CPTv}).

\begin{thebibliography}{99}
\bibitem{ciuch}
M.~Ciuchini {\it et al.}, JHEP {\bf 0308}, 031 (2002).

\bibitem{bene}
M.~Beneke {\it et al.}, Phys.\ Lett. {\bf B 576}, 173 (2003).

\bibitem{laplace}
S.~Laplace {\it et al.}, Phys.\ Rev.\ D {\bf 65}, 094040 (2002).

\bibitem{bib:KM}
 M.~Kobayashi and T.~Maskawa, Prog. Theor. Phys. {\bf 49}, 652 (1973).

\bibitem{koste}
 V.~A.~Kosteleck\`{y}, Phys.\ Rev.\ Lett. {\bf 80}, 1818 (2004).

\bibitem{yeche}
B.~Aubert {\it et al.}, The \babar\ Collaboration,
Phys.\ Rev.\ Lett. {\bf 96}, 251802 (2006).

\bibitem{babarnim}
B.~Aubert {\it et al.}, The \babar\ Collaboration, NIM A {\bf 479} (2002) 1.

\bibitem{nostra}
B.~Aubert {\it et al.}, The \babar\ Collaboration,
contributed to 33rd International Conference on High Energy Physics (ICHEP 06),
Moscow, Russia [hep-ex/0607091].

\bibitem{fmv}
B.~Aubert {\it et al.}, The \babar\ Collaboration, Phys.\ Rev.\ {\bf D}70, 
012007 (2004).

\bibitem{bozzi}
B.~Aubert {\it et al.}, The \babar\ Collaboration, Phys.\ Rev.\ Lett.\ {\bf 88}, 
231801 (2002).

\bibitem{utfit}
M.~Bona {\it et al.}, The UT{\it fit} Collaboration, hep-ph/0605213.

\end{thebibliography}
\end{document}